\documentclass{ckm}                 

\usepackage{txfonts}            

\confname{Workshop on the CKM Unitarity Triangle, IPPP Durham, April
  2003}

\title{Hadronic pollution in $B\to \rho\pi$?}

\author{AD Polosa\thanks{The author is supported by 
M. Curie fellowship, contract HPMF-CT-2001-01178}}

\address{CERN Theory Division, 1211 Geneva 23, Switzerland}

\begin{document}

\begin{abstract}

The $B\to 3\pi$ decay via $\rho$ has been shown,
in two independent calculations,  
to be very sensible to the $B\to \sigma\pi$ background.
This potentially complicates the extraction of $\alpha$ 
through the isospin analysis of $B\to \rho\pi$.
I will briefly review this topic.

\end{abstract}

\maketitle


\section{Introduction}

According to an analysis made by the E791 collaboration at
Fermilab~\cite{e791}, almost the $46\%$ of the
$D\to 3\pi$ decay rate could resonate on a scalar bump ($\sigma$)
with mass $m_\sigma=478\pm 24$~MeV and width $\Gamma_\sigma=324\pm 41$~MeV.
It is matter of discussion to assess: 1)~if this state corresponds to
a real resonance or if it is merely due to strong 
final state interactions of $\pi$'s, 2)~if
it is related to the pole on the second Riemann sheet of the 
isoscalar $S$-wave $\pi\pi$ scattering  discussed, e.g., 
in~\cite{colangelo}. In this note I will avoid further comments 
on the nature of the $\sigma$, assuming the pragmatic point of 
view that the E791 analysis indicates the presence of a state 
(or an effect) in hadronic $D$ decays
which could have some role also in $B$ decays.
In the study carried out by E791, the $\sigma$
is parameterized as a Breit-Wigner resonance; this assumption 
certainly conflicts with the broadness of the resonance one 
wants to describe,
but probably it is reasonable enough to try rough estimates 
of the  relevance of $\sigma$ in heavy meson decays.

This attitude is assumed in~\cite{prl} where a model
for heavy meson decays,~\cite{cqm}, it is used to compute the 
$D\to\sigma\pi\to 3\pi$ decay rate, tuning  the model 
parameters to fit the E791 results.
In such a way, since the model incorporates properly the spin-flavor
symmetries of HQET, one has at hand a  tool to compute
also the $B\to\sigma\pi \to 3\pi$ rate.

Why bother with such an exercise?
CLEO, BaBar and Belle~\cite{exp} measure the ratio of branching ratios:
\begin{equation}
R=\frac{{\cal B}(\bar{B}^0\to\rho^{\mp}\pi^{\pm})
}{ {\cal B}(B^-\to\rho^0\pi^-)},
\end{equation}
finding:
\begin{eqnarray}
R_{\rm CLEO}&=&2.7\pm1.3\nonumber\\
\label{eq:esperimenti}
R_{\rm BaBar}&=&1.2\pm 0.5\\
R_{\rm Belle}&=&2.6\pm 1.1\nonumber.
\end{eqnarray}
Interestingly, tree level theoretical calculations assuming 
the factorization approximation in the BSW version~\cite{bsw}
find a value of $R\sim 6$ and the inclusion of penguin contributions
does not help much in improving this result. In some other models,
like the one in Ref.~\cite{diba}, $R$ can be as large as $R\sim 13$.
A calculation including penguins in the factorization approximation
was performed in~\cite{nardulli} and found $R\sim 5.5$.

In the experimental measurements leading to $R$, 
the $\rho$ is reconstructed by cutting the
invariant mass $m_{\pi\pi}$ around the value of $m_\rho=770$~MeV
with a window $\delta\sim 200-300$~MeV.
A broad scalar bump like the $\sigma$, decaying $100\%$ in two pions,
is a good candidate to be an important background under the $\rho$.
The question here is: can the inclusion of such a background improve
the agreement of the theoretical computation of $R$ with data?

\section{Estimating the background}

To understand if the numbers could go in the right direction, 
consider the
BSW effective Hamiltonian for $B\to\sigma\pi$:
\begin{eqnarray}
H_{\rm eff}&=&\frac{G_F}{\sqrt{2}} V_{ub}^* V_{ud} \{a_1(\bar{u}b)_{V-A}
(\bar{d}u)_{V-A}\nonumber\\&+&a_2(\bar{d}b)_{V-A}(\bar{u}u)_{V-A}\},
\end{eqnarray}
where the $a_1$ term takes care of the charged $B$ decay while the $a_2$
term accounts for the neutral $B$ decay (see Fig.~1).

\begin{figure}
\hbox to\hsize{\hss
\includegraphics[width=0.4\hsize]{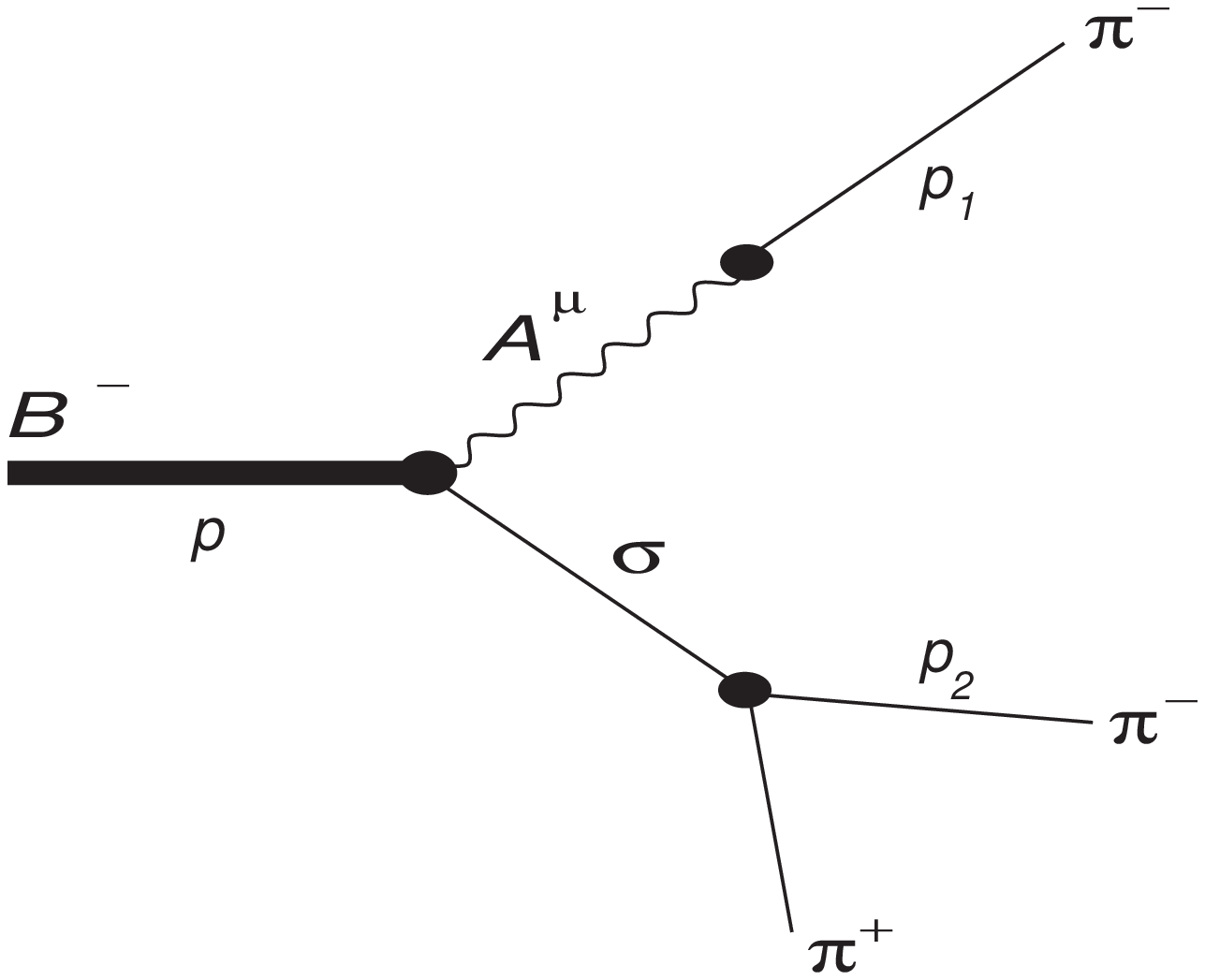}
\hss
\includegraphics[width=0.4\hsize]{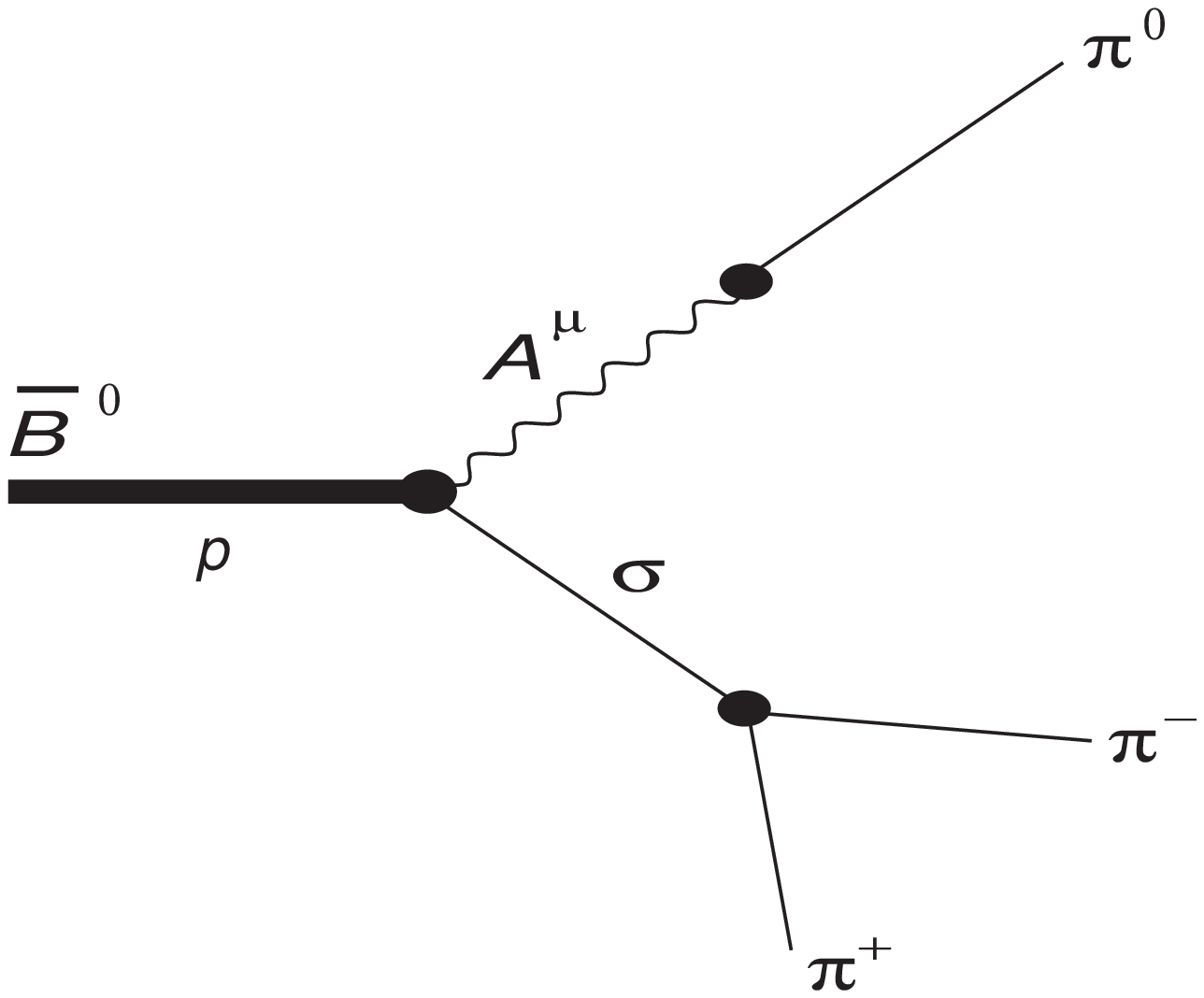}
\hss}
\vspace{-0.5truecm}
\caption{The $B\to\sigma\pi$ amplitudes mimicking the 
$B\to\rho\pi\to 3\pi$ process.}
\label{fig:otherlogos}
\end{figure}

The coefficients $a_{1,2}$ are different combination of the Wilson
coefficients fitted for $B$ decays, and it turns out that $a_1$ is
almost seven times larger than $a_2$. Therefore  the inclusion 
of the $B\to\sigma\pi$ amplitude as a background in the $(B\to 3\pi)_\rho$
decay  changes more effectively  the denominator of $R$ rather than
the numerator.
Anyway, a quantitative estimate of the effect requires the computation of 
a hadronic form factor.
The model used for this purpose is a constituent quark model based on a
Nambu-Jona-Lasinio Lagrangian whose bosonization
generates effective vertices of the kind meson$-Q-q$~\cite{cqm} 
($Q=$ heavy
quark, $q=$ light quark). A decay amplitude in this model is described by
a loop diagram with quark internal lines and meson external legs.
In our case, the computation of the hadronic form factor in the 
$D\to\sigma\pi$ amplitude requires to introduce as
a parameter the coupling $g_{\sigma q q}$ of $\sigma$ to the 
light quark component of $D$. 
This parameter can be fitted on E791 data. The same coupling enters in
the computation of $B\to\sigma\pi$.

We can estimate $g_{\sigma q q}$ by comparing the hadronic form factors
$F_0^{({\rm theor})}$ with
$F_0^{({\rm exp})}$ obtained using the E791 parameterization:
\begin{eqnarray}
&&\Gamma(D^+\to3\pi)_{{\rm via\;}\sigma}\propto \nonumber\\
&&a_1^2F_0^2g_{\pi\pi}
\int_{4m_\pi^2}^{(m_D-m_\pi)^2}dq^2 \lambda^{1/2}_{D\sigma\pi}BW(\sigma)
\lambda_{\sigma\pi\pi}^{1/2},
\end{eqnarray}
where $g_{\pi\pi}$ is the decay width of $\sigma$ in $\pi^+\pi^-$,
$\lambda_{xyz}$ is the K\"{a}ll\'{e}n triangular function defining the
two body phase space, $BW(\sigma)$ is the Breit-Wigner shape (with 
co-moving width~\cite{prl}). The comparison $F_0^{({\rm theor})}$ vs. 
$F_0^{({\rm exp})}$ for the $D\to 3\pi$ process 
gives an estimate for $g_{\sigma qq}$ which can be used to
perform the calculation of $F_0^{(B\sigma)}$ involved in the
$B\to3\pi$ process.
The hadronic form factor is defined by the matrix element:
\begin{eqnarray}
&&\langle \sigma(q_\sigma)|A^\mu(q)|H(p)\rangle=\left[ 
\frac{m_H^2-m_\sigma^2}{q^2}q^\mu \right]F_0^{(H\sigma)}(q^2)+\nonumber\\
&&\left[(p+q_\sigma)^\mu-\frac{m_H^2-m_\sigma^2}{q^2}q^\mu\right]
F_1^{(H\sigma)}(q^2),
\end{eqnarray}
where $H$ stands for heavy meson $D,B$.
Using the value obtained for $F_0^{(B\sigma)}$ in:
\begin{eqnarray}
\label{eq:b1}
\langle \sigma\pi^-|H_{\rm eff}|B^-\rangle &\propto& a_1 
F_0^{(B\sigma)}(m_\pi^2)g_{\pi\pi}f_\pi \\
\langle \sigma\pi^0|H_{\rm eff}|\bar{B}^0\rangle &\propto& a_2 
F_0^{(B\sigma)}(m_\pi^2)g_{\pi\pi}f_\pi,
\label{eq:b2}
\end{eqnarray}
we obtain a determination of the
background amplitudes to be included in the computation of $R$.
After this is done, we find $R\sim 3.5$.

The calculation of the $R$ ratio including the $\sigma$  background has
been performed in a completely different (and more refined) 
way~\cite{meissner}
avoiding the Breit-Wigner approximation (by using
a suitable scalar form factor of the pion)
and considering the $\sigma$ as the effect of strong final state pion
interactions.
The approach there used  is Chiral Perturbation Theory  
inspired. In this context it is found again 
that the impact of $\sigma$ on the
$R$ ratio is remarkably huge, and it turns out to be $R\sim 2.0-2.6$
depending on whether a $\delta$ window of $200$ or $300$~MeV is assumed
(i.e. depending on how much of the $\sigma$ tail is taken under 
the $\rho$, having in mind the picture of a $\sigma$ BW-resonance).

Another possible background under the $\rho$ could be due to $B^*$, because
the $\pi B^*B$ coupling is large. This has been investigated 
in~\cite{nardulli} and again in~\cite{tandean} with different conclusions
about its relevance.

\section{Impact on $\alpha$}

One of the methods for extracting the CKM angle $\alpha$ from data is the
isospin analysis of $B\to \rho\pi$~\cite{quinn}. The presence of a $\sigma$
background under the $\rho$ would inevitably complicate the theoretical 
analysis as long as the $\sigma$ contribution is not clearly  
isolated experimentally.

\begin{figure}
\hbox to\hsize{\hss
\includegraphics[width=\hsize]{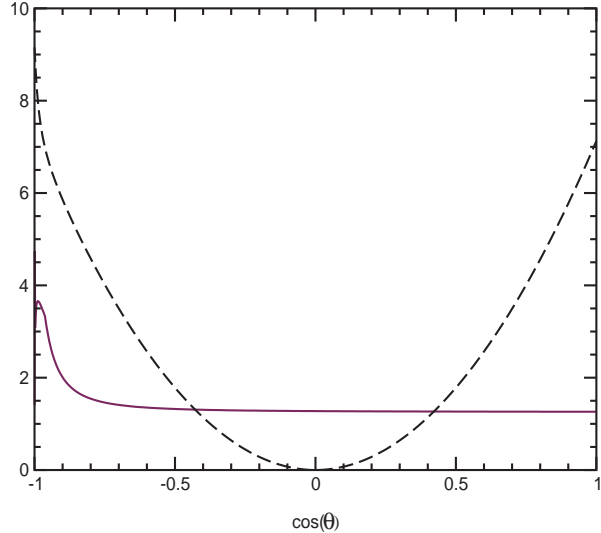}
\hss}
\caption{Matrix element squared for $B^-\to\rho^0\pi^-$ (dashed line) and
for $B^-\to\sigma\pi^-$ (solid line) as a function of 
${\rm cos}(\theta)$ where 
$\theta$ is the angle between $\pi^-$ and the $\pi^-$ from the
decay of the $\rho$ in the rest frame of the $\pi^+\pi^-$ system
generated by the $\rho$~\cite{meissner}. The difference between the two plots
reflects clearly 
the different spin nature of $\rho$ and $\sigma$.}
\end{figure}

For sure $\sigma$ and $\rho$ have quite different structures on the Dalitz 
plot (see Fig. 2 obtained in~\cite{meissner}). This means that 
in principle
one can clearly discriminate the $\sigma$ contribution experimentally.
The problem is whether the experimental sensitivity can reach the required 
level to test the possibility of the presence of a $\sigma$ in the Dalitz 
plot $B\to 3\pi$~\cite{hep-ex/0107058}. 
Even a relatively modest  $\sigma$ contribution 
(difficult to disentangle experimentally) of  
$\bar{B}^0\to\sigma\pi^0$  in $\bar{B}^0\to\rho\pi^0$
would be sufficient to spoil the isospin analysis of $B\to\rho\pi$.
In~\cite{meissner} it is shown how to enlarge the $\rho\pi$ analysis
to include the $\sigma\pi$ channel (this is possible because the
$\sigma\pi$ channel has definite properties under CP). 

The presence of this background implies that we 
gain more hadronic parameters but also more 
observables in such a way that a Dalitz plot analysis is still 
theoretically viable~\cite{meissner}.
Therefore one can conclude that if the cuts on the $\pi\pi$ invariant 
mass and on the helicity angle $\theta$ are not sufficient to suppress the
$\sigma$ in the $\rho\pi$ phase space, one is forced to consider
a more refined isospin analysis for the CKM $\alpha$ extraction.
Only more data on $B$ decays will help in deciding if this is indeed 
the case. 

A shadow on both the analyses~\cite{prl} and~\cite{meissner} is that
no one of them makes use of the improved factorization method~\cite{neubert};
a critical comparison with its predictions  
is in order~\cite{aleksan}.

Moreover the role of $\sigma$ in $D$ decays is being critically
reexamined by the
Focus collaboration following an approach quite different from the 
one adopted by E791~\cite{malvezzi}.
\section{Summary}
I have briefly reviewed the potential 
role of $\sigma$ in $D,B$ decays referring
to two distinct theoretical calculations~\cite{prl,meissner} 
of the $\sigma\pi$ background
hidden under $\rho\pi$ in $D,B\to 3\pi$ decays.
Even if the two computations stem from very different assumptions
and calculation methods, both 
point to an important impact of the $\sigma$ in $B^-\to\rho^0\pi^-$ decay.
This turns out to be helpful in understanding the experimental
numbers in Eq.~(\ref{eq:esperimenti}).

The complication of a $\sigma$ background under the $\rho$ in 
$B\to\rho\pi$ decays rises the question on how one should take it 
into account 
in the extraction of the $\alpha$ angle using the isospin analysis
of~\cite{quinn}. The data available up to now
have not yet reached the
level of sensitivity to assess if such a background would be clearly
experimentally discriminated.

I would like to thank A. Deandrea, G. Nardulli and
R. Fleisher for useful discussions.

\end{document}